\documentclass[journal,twoside]{IEEEtran}

\usepackage{epsfig,rotating,setspace,latexsym,amsmath,epsf,amssymb,amsfonts,bm,theorem}
\usepackage{cite}
\usepackage{subfigure}

\newtheorem{theorem}{Theorem}

\newtheorem{lemma}{Lemma}

\newcommand{\eqn}[1]{(\ref{#1})}
\newcommand{\fig}[1]{Fig.~\ref{#1}}
\newcommand{\mb}[1]{{\bf#1}}
\newcommand{\mathe}[1]{\hbox{E}\left[ #1 \right]}

\newcommand{\pe}{\textrm{Pr}}
\newcommand{\nl}{\nonumber\\}
\newcommand{\E}{\hbox{E}}
\newcommand{\dof}{d.o.f.~}
\newcommand{\sdof}{s.d.o.f.~}
\newcommand{\defn}{\stackrel{\triangle}{=}}
\newcommand{\bfY}{\mathbf{Y}}
\newcommand{\bfX}{\mathbf{X}}
\newcommand{\bfN}{\mathbf{N}}
\newcommand{\bfZ}{\mathbf{Z}}

\newcommand{\tN}{\tilde{N}}
\newcounter{small_constant}
\newcommand{\nextsc}{\addtocounter{small_constant}{1}  c_{\arabic{small_constant}} }
\newcommand{\nextscnu}{c_{\arabic{small_constant}}}

\newcommand{\expoints}{\textrm{Ex}}
\newcommand{\convhull}{\textrm{Co}}
\newcommand{\rank}{\textrm{rank}}

\allowdisplaybreaks

\begin{document}

\title{Secure Degrees of Freedom of Multi-user Networks: \\ One-Time-Pads in the Air via Alignment}

\author{Jianwei Xie,
        and~Sennur~Ulukus,~\IEEEmembership{Member,~IEEE}
\thanks{Manuscript received January 29, 2015,
  revised on May 7, 2015,
  and accepted June 10, 2015.
  This work was supported by NSF Grants CNS 09-64632, CCF 09-64645, CCF 10-18185 and CNS 11-47811.
}
\thanks{J. Xie was with the Department
of Electrical and Computer Engineering, University of Maryland,
College Park, MD 20742 USA. He is now with Google Inc., Mountain View, CA, 94043 USA (e-mail: xiejw@google.com)

S. Ulukus is with the Department
of Electrical and Computer Engineering, University of Maryland,
College Park, MD 20742 USA (e-mail:
ulukus@umd.edu).}}

\maketitle

\IEEEoverridecommandlockouts

\begin{abstract}
We revisit the recent secure degrees of freedom (s.d.o.f.) results for one-hop multi-user wireless networks by considering three fundamental wireless network structures: Gaussian wiretap channel with helpers, Gaussian multiple access wiretap channel, and Gaussian interference channel with secrecy constraints. We present main enabling tools and resulting communication schemes in an expository manner, along with key insights and design principles emerging from them. The main achievable schemes are based on real interference alignment, channel prefixing via cooperative jamming, and structured signalling. Real interference alignment enables aligning the cooperative jamming signals together with the message carrying signals at the eavesdroppers to protect them akin to one-time-pad protecting messages in wired systems. Real interference alignment also enables decodability at the legitimate receivers by rendering message carrying and cooperative jamming signals separable, and simultaneously aligning the cooperative jamming signals in the smallest possible sub-space. The main converse techniques are based on two key lemmas which quantify the \emph{secrecy penalty} by showing that the net effect of an eavesdropper on the system is that it eliminates one of the independent channel inputs; and the \emph{role of a helper} by developing a direct relationship between the cooperative jamming signal of a helper and the message rate. These two lemmas when applied according to the unique structure of individual networks provide tight converses. Finally, we present a blind cooperative jamming scheme for the helper network with no eavesdropper channel state information at the transmitters that achieves the same optimal \sdof as in the case of full eavesdropper channel state information.
\end{abstract}


\section{Introduction}

We consider several fundamental multi-user network structures under secrecy constraints: Gaussian wiretap channel with $M$ helpers, $K$-user Gaussian multiple access wiretap channel and $K$-user Gaussian interference channel with secrecy constraints. Security of communication was first considered by Shannon in \cite{shannon:1949}, where a legitimate pair wishes to have secure communication in the presence of an eavesdropper over a noiseless channel, leading to the necessity of secure keys and the one-time-pad encryption method, in that model. Wyner introduced the noisy wiretap channel, and demonstrated that secure communication can be attained by stochastic encoding without using any keys, if the eavesdropper is degraded with respect to the legitimate receiver \cite{wyner}. Csiszar and Korner generalized his result to arbitrary, not necessarily degraded, wiretap channels, and showed that secure communication is still possible, even when the eavesdropper is not degraded \cite{csiszar}. Csiszar and Korner introduced channel prefixing and rate splitting into the achievable scheme in addition to Wyner's stochastic encoding. Leung-Yan-Cheong and Hellman obtained the capacity-equivocation region of the Gaussian wiretap channel \cite{gaussian}, which is degraded.

This line of research has been subsequently extended to many multi-user settings, e.g., broadcast channels with confidential messages \cite{secrecy_ic, xu_bounds_bc_cm_it_09, mimo5, mimo5-1, mimo-ersen, mimo5-2}, multi-receiver wiretap channels \cite{fading1, ersen_eurasip_2009, bagherikaram_bc_2008, mimo7, mimo6}, interference channels with confidential messages and/or external eavesdroppers \cite{secrecy_ic, he_outerbound_gic_cm_ciss_09, yener-many-to-one, koyluoglu_ic_external}, multiple access wiretap channels \cite{tekin_gmac_w, cooperative_jamming, ersen_mac_allerton_08, liang_mac_cm_08, ersen_mac_book_chapter}, relay eavesdropper channels \cite{relay_1, relay_2, relay_3, relay_4}, untrusted relay channels \cite{he_untrusted_relay, ersen_crbc_2011}, two-way wiretap channels \cite{cooperative_jamming, yener-two-way, pierrot-bloch, ozan-two-way}, multi-way relay wiretap channels \cite{jorswieck}, compound wiretap channels \cite{compound_wiretap_channel, ersen_ulukus_degraded_compound}, etc. For many of these networks, even in simple Gaussian settings, exact secrecy capacity regions are still unknown. Here, we focus on Gaussian wiretap channel with helpers, Gaussian multiple access wiretap channel and Gaussian interference channel with secrecy constraints, for all of which, the exact secrecy capacity regions are unknown. In the absence of exact secrecy capacity regions, achievable secure degrees of freedom (s.d.o.f.) at high signal-to-noise ratio (SNR) regimes has been studied in the literature \cite{he_k_gic_cm_09, koyluoglu_k_user_gic_secrecy, xie_k_user_ia_compound, secrecy_ia_new, xiang_he_thesis, secrecy_ia5, raef_mac_it_12, secrecy_ia1, xie_layered_network_journal, interference_alignment_compound_channel, xie_gwch_allerton, xie_ulukus_isit_2013_mac, xie_sdof_networks_in_prepare, xie_ulukus_isit_2013_kic, xie_unified_kic, xie_asilomar_2013, xie_itw_2014, xie_sdof_region, xie_blind_cj_ciss_2013, xie_inseparability_isit_2014, khisti_arti_noise_alignment, mohamed_yener_allerton_2013, mohamed_yener_globalsip_2013}. In this paper, we revisit the results, insights, and main tools presented in a sequence of papers in \cite{xie_gwch_allerton, xie_ulukus_isit_2013_mac, xie_sdof_networks_in_prepare, xie_blind_cj_ciss_2013, xie_ulukus_isit_2013_kic, xie_unified_kic, xie_asilomar_2013, xie_itw_2014, xie_sdof_region}, which determined the exact \sdof regions of all of these three classes of networks.

\begin{figure}[t]
\centering
\includegraphics[width=0.45\textwidth]{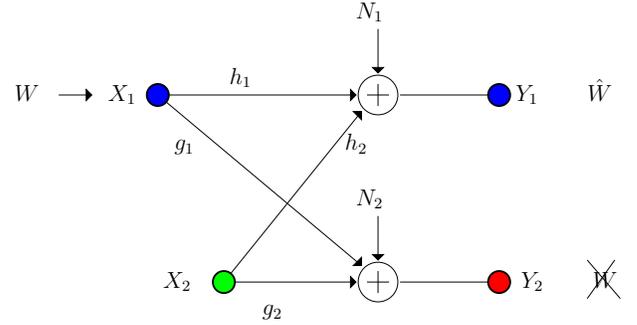}
\caption{Gaussian wiretap channel with one helper.}
\label{fig:gwc_one_helper}
\vspace*{-0.4cm}
\end{figure}

\begin{figure}[t]
\centering
\includegraphics[width=0.45\textwidth]{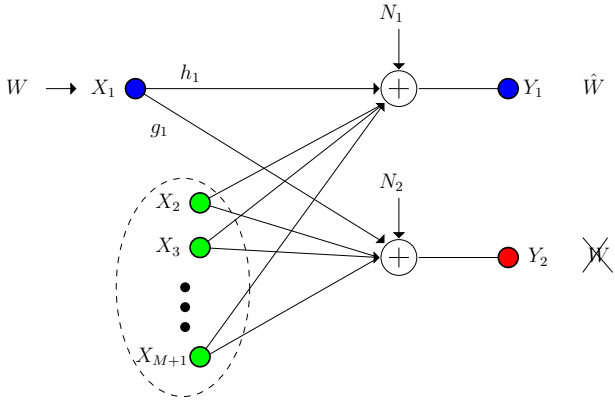}
\caption{Gaussian wiretap channel with $M$ helpers.}
\label{fig:gwc_helper_general}
\vspace*{-0.4cm}
\end{figure}

In the canonical Gaussian wiretap channel, Gaussian signalling is optimal, and the secrecy capacity is the difference of the channel capacities of the transmitter-receiver and the transmitter-eavesdropper pairs \cite{gaussian}. It is well-known that this difference does not scale with the SNR, and hence the \sdof of the Gaussian wiretap channel is zero, indicating a severe penalty due to secrecy in this case. If there is a helper in the system, as shown in \fig{fig:gwc_one_helper}, the helper can improve the achievable secrecy rate of the main transmitter by sending cooperative jamming signals \cite{tekin_gmac_w, cooperative_jamming}. The secrecy capacity of a wiretap channel with a helper, and the optimal helping strategy are unknown. However, it is known that the \sdof of this system with independent identically distributed (i.i.d.) Gaussian cooperative jamming signals is still zero \cite{wiretap_channel_with_one_helper, raef_mac_it_12}. In addition, in earlier work, it is observed that strictly positive \sdof can be obtained, for instance, by using structured codes \cite{secrecy_ia_new, xiang_he_thesis} or by using non-i.i.d.~Gaussian signalling \cite{raef_mac_it_12}. References \cite{xie_gwch_allerton, xie_sdof_networks_in_prepare} determined the exact optimal \sdof of a wiretap channel with an arbitrary number of ($M$) helpers, see Fig.~\ref{fig:gwc_helper_general}, and also the optimal helper signalling in the sense of achieving the largest \sdof The emerging idea in \cite{xie_gwch_allerton, xie_sdof_networks_in_prepare} for optimal \sdof is that the cooperation signals should not have too much randomness (hence the sub-optimality of i.i.d.~Gaussian signalling), as they hurt both the legitimate receiver and the eavesdropper. Therefore, \emph{weaker} cooperative jamming signals are needed, and that the received sub-spaces at the legitimate receiver and the eavesdropper need to be carefully controlled.

The achievable scheme in \cite{xie_gwch_allerton, xie_sdof_networks_in_prepare} is based on real interference alignment \cite{real_inter_align, real_inter_align_exploit} and cooperative jamming \cite{cooperative_jamming}, and is as follows: The legitimate receiver divides its message into $M$ parts, where $M$ is the number of helpers. Each one of $M$ helpers sends a cooperative jamming signal. All signals, both message carrying and cooperative jamming signals, come from structured pulse amplitude modulation (PAM) constellations. Each one of the cooperative jamming signals is aligned with a message carrying signal at the eavesdropper; see Fig.~\ref{fig:gwc_m_helper_ia}. This action protects the message by limiting the information leakage to the eavesdropper. This is akin to \emph{one-time-pad} in wired systems \cite{shannon:1949}. In one-time-pad, when a uniformly distributed message signal $W$ is XORed with an independent and uniformly distributed key $K$, the overall signal $X=W\oplus K$ becomes statistically independent of the message, i.e., $I(X;W)=0$, i.e., information leakage to the eavesdropper is exactly zero. With real interference alignment and uniform PAM signals, we show that the mutual information between the messages and the eavesdropper's observation is not exactly zero, but is upper bounded by a constant, and therefore, is effectively zero in terms of \sdof At the same time, all of the cooperative jamming signals are aligned in the smallest sub-space at the legitimate receiver, and are separated from the message carrying signals, see Fig.~\ref{fig:gwc_m_helper_ia}, in order to allow for the largest sub-space for the useful signals and enable their decodability. The details of the performance analysis in terms of rate and equivocation achieved by this scheme is based on the Khintchine-Groshev theorem of Diophantine approximation in number theory.

The converse developed in \cite{xie_gwch_allerton, xie_sdof_networks_in_prepare} for this channel model has two key steps. First, the secrecy rate is upper bounded by the difference of the sum of differential entropies of the channel inputs of the legitimate receiver and the helpers and the differential entropy of the eavesdropper's observation. Due to the eavesdropper's observation, one of the independent channel inputs is eliminated, and that is why this fact is named the \emph{secrecy penalty} lemma. In the second step, a direct relationship is developed between the cooperative jamming signal from an independent helper and the message rate.  The motivation of this step, which is named, \emph{role of a helper} lemma, is to determine the optimum action (role) of a helper: If the legitimate user is to reliably decode the message signal which is mixed with the cooperative jamming signal, there must exist a constraint on the cooperative jamming signal. This lemma identifies this constraint by developing an upper bound on the differential entropy of the cooperative jamming signal coming from a helper in terms of the message rate. By using these  two lemmas, and the achievable scheme described above, \cite{xie_gwch_allerton, xie_sdof_networks_in_prepare} determine the exact \sdof of the Gaussian wiretap channel with $M$ helpers to be $\frac{M}{M+1}$.

\begin{figure}[t]
\centering
\includegraphics[width=0.47\textwidth]{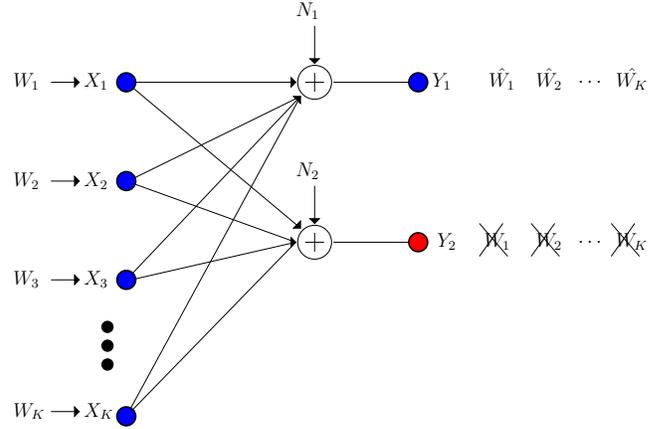}
\caption{$K$-user multiple access wiretap channel.}
\label{fig:mac_k}
\vspace*{-0.4cm}
\end{figure}

For the case of $K$-user multiple access wiretap channel, see Fig~\ref{fig:mac_k}, where all $K$ users have messages to be hidden from an external eavesdropper, \cite{xie_ulukus_isit_2013_mac,xie_sdof_networks_in_prepare} show that, the exact sum \sdof is $\frac{K(K-1)}{K(K-1)+1}$. Note that this is larger than one user utilizing the remaining $K-1$ users as helpers, which gives a \sdof of $\frac{K-1}{K}$, and time-sharing between such strategies among all users. Therefore, the fact that all users in the system have messages enables the system as a whole to obtain a higher sum \sdof The converse in this case is by extending the \emph{secrecy penalty} and \emph{role of a helper} lemmas to a multi-message setting. The achievability is by real interference alignment, channel prefixing by cooperative jamming, and structured signalling. Specifically, each transmitter divides its message into $K-1$ sub-messages, and sends these messages together with a cooperative jamming signal; see Fig~\ref{fig:mac_k_ia}. All of the signals come from the same structured PAM constellation. Each cooperative jamming signal is aligned with $K-1$ message carrying signals at the eavesdropper, protecting all of them simultaneously. At the same time, all of the $K$ cooperative jamming signals are aligned in the smallest sub-space at the legitimate receiver. Different from the helper setting, here all transmitters send a mix of message carrying signals and cooperative jamming signals. This is an instance of \emph{channel prefixing} \cite{csiszar} where the actual channel input is a further randomization of the message carrying signal.

For the case of $K$-user interference channel with secrecy constraints, \cite{xie_ulukus_isit_2013_kic, xie_unified_kic} consider the cases of \emph{confidential messages} where each transmitter's message is to be kept secret from the $K-1$ legitimate receivers, \emph{external eavesdropper} where all transmitters' messages are to be kept secret from an external eavesdropper, and the combination of the two where all messages are to be kept secret from $K$ receivers one of which is the external eavesdropper, and show that, for all of these three cases, the exact sum \sdof is $\frac{K(K-1)}{2K-1}$. Since each message is needed to be kept secret from multiple receivers, the bounding techniques in \cite{xie_sdof_networks_in_prepare} are extended in \cite{xie_ulukus_isit_2013_kic, xie_unified_kic} to be valid for the interference channel setting, by focusing on the eavesdroppers as opposed to the messages, and then by sequentially applying the \emph{role of a helper} lemma to each transmitter by treating its signal as a helper to another specific transmitter. For achievability, for the $K=2$ user interference channel with confidential messages case, since each message needs to be aligned at only two receivers, \cite{xie_sdof_networks_in_prepare} develops a real alignment and cooperative jamming based scheme as in the cases of helper and multiple access networks. However, for the general $K$-user case, each message needs to be delivered to a receiver and protected from $K$ other receivers, which requires careful simultaneous alignment at $K+1$ receivers. References \cite{xie_ulukus_isit_2013_kic, xie_unified_kic} achieve this alignment by using an \emph{asymptotical} real interference alignment technique \cite{real_inter_align_exploit}, where many signals are introduced to carry each message, and they are aligned simultaneously at multiple receivers \emph{only order-wise} (i.e., we align most of them, but not all of them), and by developing a method to upper bound the information leakage rate by a function which can be made small.

\begin{figure}[t]
\centering
\includegraphics[width=0.45\textwidth]{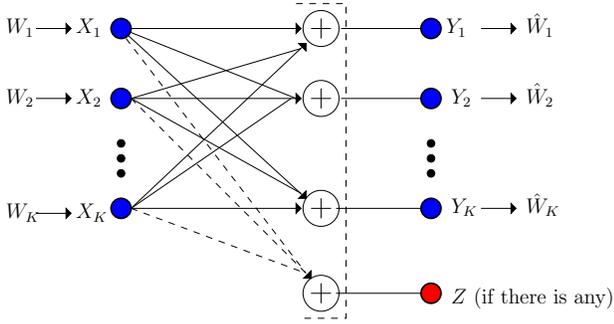}
\caption{$K$-user Gaussian interference channel with secrecy constraints.}
\label{fig:kic-general}
\vspace*{-0.4cm}
\end{figure}

While \cite{xie_ulukus_isit_2013_mac, xie_sdof_networks_in_prepare, xie_ulukus_isit_2013_kic, xie_unified_kic} determine the \emph{sum} \sdof of multiple access and interference channels with secrecy constraints, references \cite{xie_asilomar_2013, xie_itw_2014, xie_sdof_region} establish the entire \sdof \emph{regions}. Such regions show the trade-offs between the achievable \sdof of individual users. In order to determine the \sdof regions, asymmetric (not only \emph{sum}) \sdof expressions are developed. In addition, in the case of interference channels, constraints due to interference also, in addition to secrecy, are needed in the final region expressions. For achievability, \cite{xie_asilomar_2013, xie_itw_2014, xie_sdof_region} observe that the converse regions have a \emph{polytope} structure, and develop achievable schemes that achieve the extreme points of the polytope region. The major effort in \cite{xie_asilomar_2013, xie_itw_2014, xie_sdof_region} is to efficiently enumerate all of the extreme points of the converse region, and then to develop an achievable scheme for each extreme point of this region; the achievability of the entire region then follows from time-sharing.

A crucial property of all of the scenarios considered so far is that the transmitters have full channel state information (CSI) of all channels in the system. In fact, these CSI are carefully utilized in the corresponding alignment schemes. Reference \cite{xie_blind_cj_ciss_2013} considers a practically relevant scenario, where in a wiretap channel with helpers, the transmitters have CSI only to the legitimate receiver, but no CSI to the eavesdropper. Reference \cite{xie_blind_cj_ciss_2013} shows the surprising result that, in this helper network, even without any eavesdropper CSI, the optimal \sdof of $\frac{M}{M+1}$ can be achieved. The converse to this result follows from the converse for the case of full CSI in \cite{xie_gwch_allerton, xie_sdof_networks_in_prepare}. The achievability is by a \emph{blind alignment} scheme inspired by \cite{khisti_arti_noise_alignment}. In the scheme proposed in \cite{xie_blind_cj_ciss_2013}, all helpers as well as the legitimate transmitter send cooperative jamming signals; see Fig.~\ref{fig:gwc_no_csi_one_helper_no_csi_ia} and compare it with Fig.~\ref{fig:gwc_m_helper_ia}. In this system, there are a total of $M+1$ cooperative jamming signals which span the decoding space of the eavesdropper and hence protect the $M$ message carrying signals. Note that, exact alignment at the eavesdropper is not possible, as eavesdropper CSI is unknown at the transmitters. In this setting, a different technique is used to prove that the information leakage to the eavesdropper is upper bounded. In addition, here, the CSI to the legitimate receiver is used to align all of the $M+1$ cooperative jamming signals in the smallest sub-space at the legitimate receiver\footnote{Very recently, the multiple access channel \cite{pritam-isit15} and the interference channel with an external eavesdropper \cite{pritam-itw15} have been considered for the case of no eavesdropper CSI at the transmitters.}.

\section{Main Tools}
\label{sec:main_tools}

In this section, we review main tools used in this paper. The converse tools include two lemmas: Lemma~\ref{lemma:gwch_general_ub_for_m_helpers}, which is the \emph{secrecy penalty} lemma, and Lemma~\ref{lemma:gwch_general_ub_for_helper}, which is the \emph{role of a helper} lemma. The achievability tool is the technique of \emph{real interference alignment}, which is stated in Lemma~\ref{lemma:gwch_ria_real_alignment}.

\subsection{Converse Tools: Secrecy Penalty and Role of a Helper Lemmas} \label{sec:gwch_general_ub}

In the following lemma (Lemma~\ref{lemma:gwch_general_ub_for_m_helpers}), we give a general upper bound for the secrecy rate. This lemma is first motivated by, and stated for, the Gaussian wiretap channel with $M$ helpers  (see \fig{fig:gwc_helper_general}), which is defined by,
\begin{align}
\label{eqn:gwch_channel_model_helpers_genneral_1}
Y_1 &= h_1 X_1 + \sum_{j=2}^{M+1}h_j X_j + N_1 \\
\label{eqn:gwch_channel_model_helpers_genneral_2}
Y_2 & = g_1 X_1 + \sum_{j=2}^{M+1}g_j X_j + N_2
\end{align}
where $Y_1$ is the channel output of the legitimate receiver, $Y_2$ is the channel output of the eavesdropper, $X_1$ is the channel input of the legitimate transmitter, $X_i$, for $i=2,\ldots,M+1$, are the channel inputs of the $M$ helpers, $h_i$ is the channel gain of the $i$th transmitter to the legitimate receiver, $g_i$ is the channel gain of the $i$th transmitter to the eavesdropper, and $N_1$ and $N_2$ are two independent zero-mean unit-variance Gaussian random variables. All channel inputs satisfy average power constraints, $\E\left[X^2_{i}\right] \le P$, for $i=1,\ldots, M+1$. Transmitter $1$ intends to send a message $W$ to the legitimate receiver (receiver $1$). The rate of the message is $R\defn\frac{1}{n}\log|\mathcal{W}|$, where $n$ is the number of channel uses. A secrecy rate $R$ is said to be achievable if for any $\epsilon>0$ there
exists an $n$-length code such that receiver $1$ can decode this message reliably, and the message is kept information-theoretically secure against the eavesdropper,
\begin{align}
\label{eqn:secrecy_measure}
\frac{1}{n}H(W| \bfY_2) \ge \frac{1}{n} H(W) - \epsilon
\end{align}
i.e., that the uncertainty of the message $W$, given the observation $\bfY_2$ of the eavesdropper, is almost equal to the entropy of the message. This is equivalent to,
\begin{align}
\label{eqn:secrecy_measure_mi}
\frac{1}{n} I(W; \bfY_2) \le \epsilon
\end{align}
i.e., the (normalized) information leakage to the eavesdropper asymptotically vanishes, resulting in perfect (weak) secrecy \cite{csiszar}. The supremum of all achievable secrecy rates is the secrecy capacity $C_s$, and the
s.d.o.f., $D_s$, is defined as
\begin{align}
D_s \defn \lim_{P\to\infty}  \frac{C_s}{\frac{1}{2}\log P}
\label{eqn:sec-dof-defn}
\end{align}
The s.d.o.f.~determines the scaling of the secrecy capacity with the capacity of a single-user channel which is $\frac{1}{2}\log P$ at high SNR. That is, s.d.o.f.~is the pre-log factor of the secrecy capacity at high SNR.

The goal of Lemma~\ref{lemma:gwch_general_ub_for_m_helpers} is to quantify the \emph{secrecy penalty} due to the presence of an eavesdropper. We work with $n$-letter signals (hence bold vectors) and introduce small independent Gaussian fudge variables $\tilde{N}_i$ and state inequalities in terms of slightly perturbed channel inputs $\tilde{X}_i$; this is for regularity purposes only, so that we can use differential entropies even for discrete signals throughout the paper.

This lemma states that the secrecy rate of the legitimate pair is upper bounded by the difference of the sum of differential entropies of all channel inputs (perturbed by small noise) and the differential entropy of the eavesdropper's observation; see \eqn{eqn:gwch_general_ub_for_m_helpers}. This upper bound can  be interpreted as follows: If we consider the eavesdropper's observation as the \emph{secrecy penalty,} then the secrecy penalty is
tantamount to the elimination of one of the channel inputs in the system; see \eqn{eqn:gwch_general_ub_for_m_helpers_kill_y2}.

\begin{lemma}[\!\! \cite{xie_gwch_allerton, xie_sdof_networks_in_prepare}] \label{lemma:gwch_general_ub_for_m_helpers}
[Secrecy penalty lemma] The secrecy rate of the legitimate pair is upper bounded as
\begin{align}
n R
& \le \sum_{i=1}^{M+1} h( \tilde\bfX_i)  - h(\bfY_2) +  n c \label{eqn:gwch_general_ub_for_m_helpers} \\
& \le \sum_{i=1,i\neq j}^{M+1} h( \tilde\bfX_i) +  n c' \label{eqn:gwch_general_ub_for_m_helpers_kill_y2}
\end{align}
where  $\tilde\bfX_i = \bfX_i+\tilde\bfN_i$ for $i=1,2,\cdots,M+1$, and $\tilde\bfN_i$ is an i.i.d.~sequence (in time) of random variables $\tilde N_i$ which are independent Gaussian random variables with zero-mean and variance  $\tilde\sigma_{i}^2$ with $\tilde\sigma_{i}^2<\min(1/h_i^2,1/g_i^2)$. In addition, $c$ and $c'$ are constants which do not depend on $P$, and $j \in \{1,2,\cdots,M+1\}$ could be arbitrary.
\end{lemma}

In the following lemma (Lemma~\ref{lemma:gwch_general_ub_for_helper}), we give a general upper bound for the differential entropy of the signal of a helper based on the decodability of the message of the legitimate transmitter at the legitimate receiver. This lemma is also motivated in the helper setting, but as with Lemma~\ref{lemma:gwch_general_ub_for_m_helpers} above, it is valid for more general settings. The goal of this lemma is to quantify the \emph{role of a helper}, in terms of its affect on the system. In this lemma, $W$ is the message of the legitimate transmitter, and its entropy $H(W)$ is the message rate. Here, $X_j$ is the $j$th helper's channel input, and $Y_1$ is the legitimate receiver's channel output. Again, we use slightly perturbed channel inputs for regularity.

This lemma is motivated as follows: A cooperative jamming signal from a helper may potentially increase the secrecy of the legitimate transmitter-receiver pair by creating extra equivocation at the eavesdropper. However, if the helper creates too much equivocation, it may also hurt the decoding performance of the legitimate receiver. Since the legitimate receiver needs to decode message $W$ by observing $Y_1$, there must exist a constraint on the cooperative jamming signal of the helper, $X_j$. This lemma develops a constraint on the differential entropy of (the noisy version of) the cooperative jamming signal of any given helper, helper $j$ in \eqn{eqn:gwch_general_ub_for_helper}, in terms of the differential entropy of the legitimate user's channel output and the message rate $H(W)$. The inequality in \eqn{eqn:gwch_general_ub_for_helper} states that, for a given message rate $H(W)$, the entropy of the signal that the helper puts into the channel should not be too much. Alternatively, $H(W)$ can be moved to the left hand side of \eqn{eqn:gwch_general_ub_for_helper}, and this inequality can be interpreted as an upper on the message rate given the helper signal's entropy. In particular, the higher the differential entropy of the cooperative jamming signal the lower this upper bound on the message rate will be. This motivates us not to use i.i.d.~Gaussian cooperative jamming signals which have the highest differential entropy.

\begin{lemma}[\!\! \cite{xie_gwch_allerton, xie_sdof_networks_in_prepare}] \label{lemma:gwch_general_ub_for_helper}
[Role of a helper lemma] For reliable decoding at the legitimate receiver, the differential entropy of the input signal of helper $j$, $\bfX_j$, must satisfy
\begin{equation}
 h(\bfX_j + \tilde\bfN)\le  h(\bfY_1)  - H(W) + n {c}
  \label{eqn:gwch_general_ub_for_helper}
\end{equation}
where $c$ is a constant which does not depend on $P$, and $\tN$ is a new Gaussian noise independent of all other random variables with $\sigma_{\tN}^2 < \frac{1}{h_j^2}$, and $\tilde\bfN$ is an i.i.d.~sequence of $\tilde N$.
\end{lemma}

\subsection{Achievability Tools: Real Interference Alignment}

In this subsection, we review pulse amplitude modulation (PAM) and real interference alignment \cite{real_inter_align_exploit, real_inter_align}, similar to the review in \cite[Section~III]{interference_alignment_compound_channel}. The purpose of this subsection is to illustrate that by using real interference alignment, the transmission rate of a PAM scheme can be made to approach the Shannon achievable rate at high SNR. This provides a universal and convenient way to design capacity-achieving signalling schemes at high SNR by using PAM for different channel models as will be done in later sections.

For a point-to-point scalar Gaussian channel,
\begin{equation}
Y = X + Z
\end{equation}
with additive Gaussian noise $Z$ of zero-mean and variance $\sigma^2$, and an input power constraint $\mathe{X^2} \le P$, assume that the input symbols are drawn from a PAM constellation,
\begin{equation}
C(a,Q) = a \left\{ -Q, -Q+1, \ldots, Q-1,Q\right\}
\label{constel}
\end{equation}
where $Q$ is a positive integer and $a$ is a real number to normalize the transmit power. Note that, $a$ is also the minimum distance $d_{min}(C)$ of this constellation, which has the probability
of error
\begin{equation}
\pe(e) = \pe\left[ X \neq \hat X\right] \le \exp\left(  - \frac{d_{min}^2}{8 \sigma^2}\right)  = \exp\left(  - \frac{a^2}{8 \sigma^2}\right)
\end{equation}
where $\hat X$  is an estimate for $X$ obtained by choosing the closest point in the constellation $C(a,Q) $ based on observation $Y$.

This PAM scheme for the point-to-point scalar channel can be generalized to multiple data streams. Let the transmit signal be
\begin{equation}
x = \mb{a}^T \mb{b} = \sum^L_{i=1} a_i b_i
\end{equation}
where $a_1,\ldots, a_L$ are rationally independent real numbers\footnote{ $a_1, \ldots, a_L$ are rationally independent if whenever $q_1,\ldots,q_L$ are rational numbers then $\sum^L_{i=1} q_i a_i =0$  implies $q_i=0$ for all $i$.} and each $b_i$ is drawn independently from the constellation $C(a,Q)$ in (\ref{constel}). The real value $x$ is a combination of  $L$ data streams, and the constellation observed at the receiver consists of $(2 Q+1)^L$ signal points.

By using the Khintchine-Groshev theorem of Diophantine approximation in number theory, \cite{real_inter_align_exploit,real_inter_align} bounded the minimum distance $d_{min}$ of points in the receiver's constellation: For any $\delta>0$, there exists a constant $k_\delta$, such that
\begin{equation}
\label{ria:lb_of_d}
d_{min} \ge \frac{ k_\delta  a}{Q^{L-1+\delta}}
\end{equation}
for almost all rationally independent $\{a_i\}_{i=1}^L$, except for a set of Lebesgue measure zero. Since the minimum distance of the receiver constellation is lower bounded, with proper choice of $a$ and $Q$, the probability of error can be made arbitrarily small, with rate $R$ approaching $\frac{1}{2} \log P$.  This result is stated in the following lemma, as in \cite[Proposition~3]{interference_alignment_compound_channel}.

\begin{lemma}[\!\! \cite{real_inter_align_exploit,real_inter_align}] \label{lemma:gwch_ria_real_alignment}
[Real interference alignment] For any small enough $\delta>0$, there exists a positive constant
$\gamma$, which is independent of $P$, such that if we choose
\begin{equation}
Q = P^{\frac{1-\delta}{2(L+\delta)}}
\qquad \mbox{and} \qquad
a=\gamma \frac{P^{\frac{1}{2}}}{Q}
\end{equation}
then the average power constraint is satisfied, i.e., $\mathe{X^2}\le P $, and for almost all $\{a_i\}_{i=1}^L$, except for a set of Lebesgue measure zero, the probability of error is bounded by
\begin{equation}
  \mathrm{Pr}(e) \le \exp\left( - \eta_\gamma P^{{ \delta}} \right)
\end{equation}
where $\eta_\gamma$ is a positive constant which is independent of $P$.
\end{lemma}

\section{Wiretap Channels with $M$ Helpers} \label{sec:gwch_m_helper}

In this section, we consider the Gaussian wiretap channel with $M$ helpers shown in Fig.~\ref{fig:gwc_helper_general} and defined in \eqn{eqn:gwch_channel_model_helpers_genneral_1} and \eqn{eqn:gwch_channel_model_helpers_genneral_2}.

In the sequel, we will demonstrate the use of converse and achievability lemmas presented in Section~\ref{sec:main_tools} in some depth in the context of a helper network; we will then make much briefer presentations for the multiple access and interference networks in the following sections.

Here, we show that for the wiretap channel with $M$ helpers, the exact \sdof is $\frac{M}{M+1}$, as stated in the following theorem. This shows that even though the helpers are independent, the \sdof increases monotonically with the number of helpers $M$, and goes to 1, which is the \dof with no secrecy constraints.

\begin{theorem}[\!\!\cite{xie_gwch_allerton, xie_sdof_networks_in_prepare}]
The \sdof of the Gaussian wiretap channel with $M$ helpers is $\frac{M}{M+1}$ for almost all channel gains.
\end{theorem}

\subsection{Converse}
We start with \eqn{eqn:gwch_general_ub_for_m_helpers_kill_y2} of Lemma~\ref{lemma:gwch_general_ub_for_m_helpers} with the selection of $j=1$
\begin{align}
n R
& \le\sum_{i=1,i\neq j}^{M+1} h( \tilde\bfX_i) +  n c' \\
& = \sum_{i=2}^{M+1} h( \tilde\bfX_i) +  n c' \\
& \le  M [ h(\bfY_1) -H(W)]  +  n \nextsc
\label{eqn:proof_m_helpers_next}
\end{align}
where \eqn{eqn:proof_m_helpers_next} is due to Lemma~\ref{lemma:gwch_general_ub_for_helper} for each cooperative jamming signal $\tilde\bfX_i$, $i=2,\cdots,M+1$. By noting $H(W)=nR$, \eqn{eqn:proof_m_helpers_next} implies that
\begin{align}
(M+1) n R & \le  M h(\bfY_1)  + n \nextscnu \label{eqn:gwch_ub_for_m_helpers} \\
& \le M \left( \frac{n}{2} \log P \right) + n \nextsc
\end{align}
which further implies  that
\begin{equation}
D_s \le \frac{M}{M+1}
\end{equation}
which concludes the converse part of the theorem.

\begin{figure*}[t]
\centering
\includegraphics[width=0.8\textwidth]{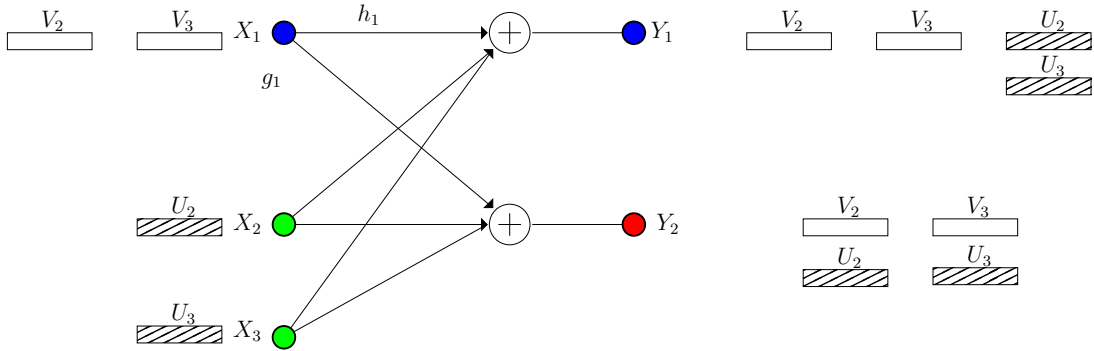}
\caption{Illustration of interference alignment for the Gaussian wiretap channel with M helpers. Here, $M=2$.}
\label{fig:gwc_m_helper_ia}
\vspace*{-0.4cm}
\end{figure*}

\subsection{Achievable Scheme}

Let $\{V_2,V_3,\cdots,V_{M+1},U_2,U_3,\cdots,U_{M+1}\}$ be mutually independent discrete random variables, each of which uniformly drawn from the same PAM constellation $C(a,Q)$ in (\ref{constel}), where $a$ and $Q$ will be specified later. We choose the input signal of the legitimate transmitter as
\begin{equation}
X_1  = \sum_{k=2}^{M+1} \frac{g_k}{g_1 h_k}  V_k
\end{equation}
and the input signal of the $j$th helper, $j=2,\cdots,M+1$, as
\begin{equation}
X_j = \frac{1}{h_j} U_j
\end{equation}
Then, the observations of the receivers are
\begin{align}
Y_1 & = \sum_{k=2}^{M+1} \frac{h_1 g_k}{g_1 h_k} V_k+ \left[ \sum_{j=2}^{M+1} U_j \right]+ N_1 \label{helper-align-su1}\\
Y_2& = \sum_{k=2}^{M+1} \frac{g_k}{ h_k} \Big[ V_k + U_k \Big]+ N_2 \label{helper-align-su2}
\end{align}

The intuition here is as follows: We use $M$ independent sub-signals $V_k$, $k=2,\cdots,M+1$, to represent the signals carrying the original message $W$. The input signal $X_1$ is a linear combination of $V_k$s. To cooperatively jam the eavesdropper, each helper $k$ aligns the cooperative jamming signal $U_k$ in the same \emph{dimension} as the sub-signal $V_k$ at the eavesdropper. At the legitimate receiver, all of the cooperative jamming signals $U_k$s are well-aligned such that they occupy a  small portion of the signal space. Since, with probability one, $\left\{1,\frac{h_1 g_2}{g_1 h_2}, \frac{h_1 g_3}{g_1 h_3}, \cdots, \frac{h_1 g_{M+1}}{g_1 h_{M+1}}\right\}$ are rationally independent, signals $\left\{V_2,V_3,\cdots,V_{M+1}, \sum_{j=2}^{M+1} U_j \right\}$ can be distinguished by the legitimate receiver. Square parentheses in (\ref{helper-align-su1}) and (\ref{helper-align-su2}) indicate alignments at the two receivers. As an example, the case of $M=2$ is shown in Fig.~\ref{fig:gwc_m_helper_ia}.

The exact performance analysis supporting the above intuition is based on real interference alignment summarized in Lemma~\ref{lemma:gwch_ria_real_alignment}, and the achievable secrecy rate in \cite{csiszar}. In particular, since, for each $j\neq 1$, ${\bfX}_j$ is an i.i.d.~sequence and independent of ${\bfX}_1$, the following secrecy rate is achievable \cite{csiszar}
\begin{equation}
C_s \ge I(X_1;Y_1) - I(X_1;Y_2)
\label{ck-formula}
\end{equation}

Now, we first bound the probability of decoding error. Note that the \emph{space} observed at receiver $1$ consists of $(2Q+1)^M (2MQ+1)$ points in $M+1$ \emph{dimensions}, and the sub-signal in each \emph{dimension} is drawn from a constellation of $C(a,MQ)$. Here, we use the property that $C(a,Q)\subset C(a,MQ)$.  By Lemma \ref{lemma:gwch_ria_real_alignment}, for any small enough $\delta>0$ and for almost all rationally independent $\left\{1,\frac{h_1 g_2}{g_1 h_2}, \frac{h_1 g_3}{g_1 h_3}, \cdots, \frac{h_1 g_{M+1}}{g_1 h_{M+1}}\right\}$, except for a set of Lebesgue measure zero, there exists a positive constant $\gamma$, which is independent of $P$, such that if we choose $Q = P^{\frac{1-\delta}{2(M+1+\delta)}}$ and $a=\gamma P^{\frac{1}{2}}/Q$ then the average power constraint is satisfied and the probability of error is bounded as
\begin{equation}
  \pe\left[X_1\neq\hat{X}_1\right]
\le
\exp\left( - \eta_\gamma P^{{ \delta}} \right)
\end{equation}
where $\eta_\gamma$ is a positive constant which is independent of $P$ and where $\hat{X}_1$ is the estimate of $X_1$ by choosing the closest point in the constellation based on observation $Y_1$. This shows that the legitimate receiver can decode the messages reliably.

By Fano's inequality and the Markov chain $X_1\rightarrow Y_1\rightarrow\hat{X}_1$, we know that
\begin{align}
H(X_1 | Y_1)
& \le H(X_1|\hat{X}_1) \\
& \le 1 +
\exp\left( - \eta_\gamma P^{{ \delta}}
\right) \log(2Q+1)^M
\end{align}
which means that
\begin{align}
I(X_1;Y_1) & = H(X_1) - H(X_1|Y_1) \\
& \ge
\left[ 1-
\exp\left( \eta_\gamma P^{{ \delta}}
\right) \right] \log(2Q+1)^M  -1
\label{eqn:gwch_wiretap_m_helper_lb_ixy1}
\end{align}
On the other hand,
\begin{align}
  I(X_1;Y_2)
& \le I\left(X_1;\sum_{k=2}^{M+1} \frac{g_k}{ h_k} ( V_k + U_k )\right) \\
&  =  H\left( \sum_{k=2}^{M+1} \frac{g_k}{ h_k} ( V_k + U_k )\right) \nl
&\quad
- H\left( \sum_{k=2}^{M+1} \frac{g_k}{ h_k} ( V_k + U_k )\Big|X_1\right) \\
&  =  H\left( \sum_{k=2}^{M+1} \frac{g_k}{ h_k} ( V_k + U_k )\right)- H\left( \sum_{k=2}^{M+1} \frac{g_k}{ h_k} U_k\right) \\
& \le \log( 4Q +1)^M - \log( 2Q +1 )^M
\label{eqn:gwch_ub_for_combined_constellation_m_streams} \\
& \le M \log\frac{ 4Q +1 } { 2Q +1 } \\
& \le M
\label{eqn:gwch_wiretap_m_helper_lb_ixy2}
\end{align}
where \eqn{eqn:gwch_ub_for_combined_constellation_m_streams} is due to the fact that entropy of the sum $\sum_{k=2}^{M+1} \frac{g_k}{ h_k} ( V_k + U_k )$ is maximized by the uniform distribution which takes values over a set of cardinality $(4Q +1)^M$.

Combining \eqn{eqn:gwch_wiretap_m_helper_lb_ixy1} and \eqn{eqn:gwch_wiretap_m_helper_lb_ixy2}, from (\ref{ck-formula}), we have
\begin{align}
C_s
& \ge I(X_1;Y_1) - I(X_1;Y_2) \\
& \ge \left[ 1-\exp\left(  -\eta_\gamma P^{{ \delta}}
\right) \right] \log(2Q+1)^M  -(M+1) \\
& \ge \left[ 1-\exp\left(  -\eta_\gamma P^{{ \delta}}
\right) \right] \log(2P^{\frac{1-\delta}{2(M+1+\delta)}}+1)^M  -(M+1) \\
& ={\frac{M(1-\delta)}{(M+1+\delta)}} \left(\frac{1}{2}\log P\right) + o(\log P)
\end{align}
where $o(\cdot)$ is the little-$o$ function. If we choose $\delta$ arbitrarily small, then we can achieve $\frac{M}{M+1}$ s.d.o.f., which concludes the achievability part of the theorem.

\section{Multiple Access Wiretap Channel} \label{sec:gmacw_k}

In this section, we consider the $K$-user multiple access wiretap channel shown in Fig.~\ref{fig:mac_k}, which has multiple transmitters each with its own message to transmit:
\begin{align}
\displaystyle Y_1 & = \sum_{i=1}^{K} h_i X_i + N_1 \\
\displaystyle Y_2 & = \sum_{i=1}^{K} g_i X_i + N_2
\end{align}
We show that the exact sum \sdof of this channel is $\frac{K(K-1)}{K(K-1)+1}$, as stated in the following theorem. Note that this is strictly larger than the \sdof of the corresponding helper network, which is $\frac{K-1}{K}$.

\begin{theorem}[\!\!\cite{xie_ulukus_isit_2013_mac, xie_sdof_networks_in_prepare}]
The sum \sdof of the $K$-user Gaussian multiple access wiretap channel is $\frac{K(K-1)}{K(K-1)+1}$ for almost all channel gains.
\end{theorem}

The converse is derived by starting with an upper bound which is similar to the \emph{secrecy penalty} lemma in Lemma~\ref{lemma:gwch_general_ub_for_m_helpers}, and considering all transmitters as a single virtual transmitter:
\begin{align}
n \sum_{i=1}^K R_i
& \le \sum_{i=1}^Kh(\tilde\bfX_i)   -h(\bfY_2) + n \nextsc \\
& \le \sum_{i=2}^Kh(\tilde\bfX_i) + n \nextsc
\label{eqn:mac_k_ub_y_2_by_x_1}
\end{align}
In addition, similar to the \emph{role of a helper} lemma in Lemma~\ref{lemma:gwch_general_ub_for_helper}, we bound the differential entropy of each user's channel input with the sum of the decodable rates of all other users:
\begin{align}
\sum_{i\neq j} H( W_{i})
& =   H( W_{\neq j})  \le h( \bfY_1) - h(\tilde\bfX_j) + n \nextsc
\end{align}
The converse is completed by proceeding similarly to the case of the helper network, starting from the above generalizations of Lemmas~\ref{lemma:gwch_general_ub_for_m_helpers} and \ref{lemma:gwch_general_ub_for_helper}.

\begin{figure}[t]
\centering
\includegraphics[width=0.5\textwidth]{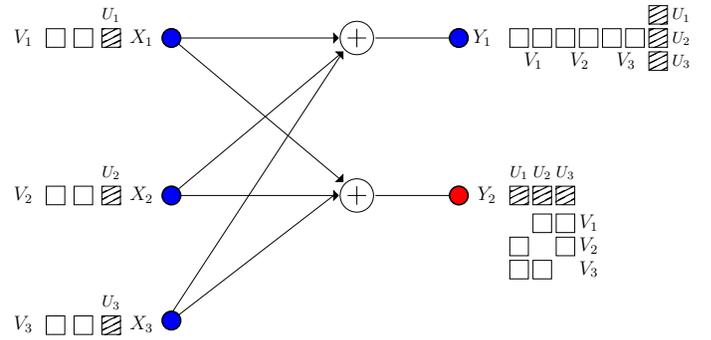}
\caption{Illustration of interference alignment for the $K$-user multiple
access wiretap channel. Here, $K=3$.}
\label{fig:mac_k_ia}
\vspace*{-0.4cm}
\end{figure}

The achievable scheme is as follows: Each transmitter $i$ divides its message into $K-1$ mutually independent sub-signals. In addition, each transmitter $i$ sends a cooperative jamming signal $U_i$. This is an instance of \emph{channel prefixing} \cite{csiszar}, where the channel input is further randomized. At the eavesdropper $Y_2$, each sub-signal indexed by $(i,j)$, where $j\in\{1,\cdots,K\}\backslash \{i\}$, is \emph{aligned} with a cooperative jamming signal $U_i$. At the legitimate receiver $Y_1$, all of the cooperative jamming signals are \emph{aligned} in the same dimension to \emph{occupy} as \emph{small} a signal space as possible. This scheme is illustrated in Fig.~\ref{fig:mac_k_ia} for the case of $K=3$.

Specifically, we use in total $K^2$ mutually independent random variables which are
\begin{align}
& V_{i,j},\quad i,j\in \{1,2,\cdots,K\}, j\neq i \\
& U_k ,\quad k\in\{1,2,\cdots,K\}
\end{align}
where $V_{i,j}$, $j \neq i$ are the $K-1$ sub-signals that carry the message of user $i$, and $U_i$ is the cooperative jamming signal sent by user $i$. All of these random variables are uniformly and independently drawn from the same constellation $C(a,Q)$ in (\ref{constel}). For each $i\in\{1,2,\cdots,K\}$, we choose the input signal of transmitter $i$ as
\begin{equation}
X_i  = \sum_{j=1,j\neq i}^{K} \frac{g_j}{g_i h_j}  V_{i,j} + \frac{1}{h_i} U_i
\end{equation}
With these input signal selections, received signals are
\begin{align}
Y_1& = \sum_{i=1}^K \sum_{j=1,j\neq i}^{K} \frac{g_j h_i}{g_i h_j}  V_{i,j}
   + \left[ \sum_{k=1}^K U_k \right] + N_1 \label{align-su1} \\
Y_2& =\left( \sum_{i=1}^K \sum_{j=1,j\neq i}^{K} \frac{g_j} { h_j}  V_{i,j} \right)
   + \sum_{j=1}^K \frac{g_j}{h_j} U_j+ N_2 \\
& = \sum_{j=1}^K\frac{g_j}{h_j}\left[U_j +\sum_{i=1,i\neq j}^{K}  V_{i,j}\right]+ N_2 \label{align-su2}
\end{align}
Each of the signals in the square parentheses in (\ref{align-su1}) and (\ref{align-su2}) are \emph{aligned} in the same irrational \emph{dimension}. This alignment in (\ref{align-su1}) ensures that the cooperative jamming signals occupy the smallest possible space at the legitimate receiver, and the alignment in (\ref{align-su2}) ensures that each $U_j$ protects all the $V_{i,j}$s in the same square parentheses.

\section{Interference Channel with Secrecy} \label{sec:kic_secrecy}

In this section, we consider the $K$-user Gaussian interference channel with secrecy constraints shown in Fig.~\ref{fig:kic-general}. The channel model is:
\begin{align}
  \label{eqn:kic-channel-model-ee-1}
  Y_i & = \sum_{j=1}^K h_{ji} X_j + N_i, \qquad i =1,\ldots,K \\
  \label{eqn:kic-channel-model-ee-2}
  Z   & = \sum_{j=1}^K g_{j} X_j + N_Z \qquad \mbox{(if there is any)}
\end{align}
which has not only  multiple transmitters but also multiple receivers in the network. We consider three different secrecy requirements: interference channel with an external eavesdropper (IC-EE), where all of the messages are kept secure against the external eavesdropper; interference channel with confidential messages (IC-CM), where all messages are kept secure against unintended receivers; and their combination (IC-CM-EE), where all messages are kept secure against all unintended receivers and the eavesdropper. The sum \sdof is the same for all three networks and is stated in the following theorem.

\begin{theorem}[\!\! \cite{xie_ulukus_isit_2013_kic,xie_unified_kic}]
\label{eqn:kic-ds-final}
The sum \sdof of the $K$-user IC-EE, IC-CM, and IC-CM-EE is $\frac{K(K-1)}{2K-1}$ for almost all channel gains.
\end{theorem}

We provide an outline of the converse and achievable scheme for IC-EE only here. The converse starts with Lemma~\ref{lemma:gwch_general_ub_for_m_helpers}, the \emph{secrecy penalty} lemma: For any $j=1,\ldots,K$,
\begin{align}
n\sum_{i=1}^K R_i
& \le h(\tilde\bfX_1^K) -h(\bfZ) + n c_3 \\
& \le \sum_{i=1}^K h(\tilde\bfX_i) -h(\bfZ) + n c_3\\
& \le \sum_{i=1,i\neq j}^K h(\tilde\bfX_i) + n \nextsc \label{apply-rohl}
\end{align}
Then, we apply the \emph{role of a helper} lemma, Lemma~\ref{lemma:gwch_general_ub_for_helper}, to  each $\tilde\bfX_i$ with $k=i+1$ (for $i=K$, $k=1$), in (\ref{apply-rohl}) as
\begin{align}
n\sum_{i=1}^K R_i
& \le h(\tilde\bfX_1) +  h(\tilde\bfX_2) + \cdots
+ h(\tilde\bfX_{j-1}) \nl
& \quad +h(\tilde\bfX_{j+1}) + \cdots + h(\tilde\bfX_K) + n\nextsc
\\
& \le \left[ h(\bfY_2) - n R_2 \right]
+ \left[ h(\bfY_3) - n R_3 \right] + \cdots \nl
& \quad + \left[h(\bfY_{j}) - n R_{j} \right] + \left[ h(\bfY_{j+2}) - n R_{j+2} \right]  + \cdots \nl
&\quad
 +\left[ h(\bfY_{K}) - n R_{K} \right] + \left[ h(\bfY_1) - n R_1 \right]
+ n \nextsc
\end{align}
By noting that $h(\bfY_i) \le \frac{n}{2}\log P + n c_i'$ for each $i$, we have
\begin{align}
2 n\sum_{i=1}^K R_i \le (K-1) \left( \frac{n}{2}\log P \right) + n R_{{(j+1)} \bmod{K}} + n \nextsc
\label{eqn:kic-ee-ub-general-j}
\end{align}
for $j=1,\ldots,K$. Therefore, we have a total of $K$ bounds in \eqn{eqn:kic-ee-ub-general-j} for $j=1,\ldots,K$. Summing these $K$ bounds, we obtain:
\begin{align}
(2 K - 1) n\sum_{i=1}^K R_i \le K (K-1) \left( \frac{n}{2}\log P \right) + n \nextsc
\end{align}
which gives
\begin{align}
  D_{s,\Sigma} \le \frac{ K (K-1)}{2 K - 1}
\end{align}
completing the converse for IC-EE.

\begin{figure}[t]
\centering
\includegraphics[width=0.45\textwidth]{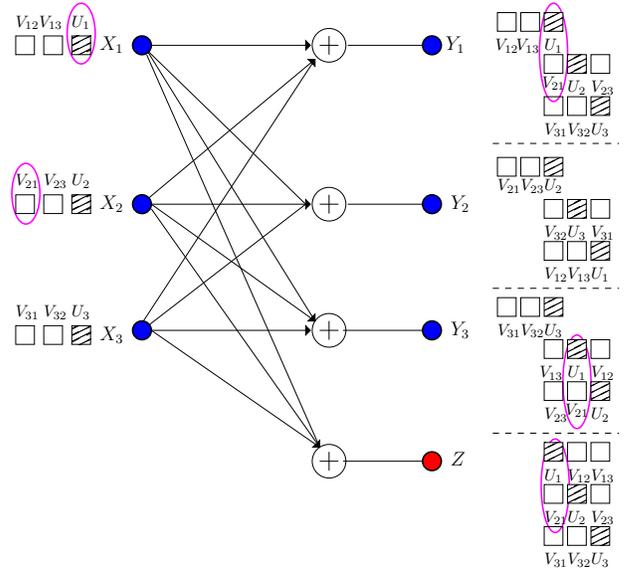}
\caption{Illustration of  alignment for $3$-user
IC-CM-EE. $U_1$ and $V_{21}$ are marked to emphasize their
simultaneous alignment at $Y_1$, $Y_3$ and $Z$. }
\label{fig:kic_alignment}
\vspace*{-0.4cm}
\end{figure}

The achievability is based on Lemma~\ref{lemma:gwch_ria_real_alignment} for the $K$-user IC-CM-EE, which will imply achievability for $K$-user IC-EE. We employ a total of $K^2$ random variables,
\begin{align}
 V_{ij}, \quad & i,j = 1,\ldots,K, \, j\neq i \\
 U_k, \quad &k = 1,\ldots,K
\end{align}
which are illustrated in Fig.~\ref{fig:kic_alignment} for the case of $K=3$. For transmitter $i$, $K-1$ random
variables $\{V_{ij}\}_{j\neq i}$, each representing a sub-message, collectively carry message $W_i$. Different than before, rather than protecting one message at one receiver, each $U_k$ simultaneously protects a portion of all sub-messages at all required receivers. More specifically, $U_k$ protects  $\{V_{ik}\}_{i\neq k,i\neq j}$ at receivers $j$, and at the eavesdropper (if there is any). For example, in Fig.~\ref{fig:kic_alignment}, $U_1$ protects $V_{21}$ and $V_{31}$ where necessary. In particular, $U_1$ protects $V_{21}$ at receivers 1, 3 and the eavesdropper; and it protects $V_{31}$ at receivers 1, 2 and the eavesdropper. As a technical challenge, this requires $U_1$ to be aligned with the same signal, say $V_{21}$, at multiple receivers simultaneously, i.e., at receivers 1, 3 and the eavesdropper. These particular alignments are circled by ellipsoids in Fig.~\ref{fig:kic_alignment}. We do these simultaneous alignments using asymptotic real alignment technique proposed in \cite{real_inter_align_exploit} and used in \cite{xie_k_user_ia_compound, interference_alignment_compound_channel}.

\section{S.d.o.f. Regions of Wireless Networks}

In this section, we revisit the $K$-user multiple access wiretap channel in Section~\ref{sec:gmacw_k} and $K$-user interference channel in Section~\ref{sec:kic_secrecy}, and study the \sdof \emph{regions} of both networks. The results have been characterized in the following theorems.

\begin{theorem}[\!\! \cite{xie_asilomar_2013,xie_sdof_region}]
\label{theo:sdof_mac_capacity_region}
The \sdof region $D$ of the $K$-user multiple access wiretap channel
is the set of all $\mathbf{d}$ satisfying
\begin{align}
   K d_i + (K-1) \sum_{j=1,j\neq i}^K d_j  &\le K-1, & i=1,\ldots,K
\label{eqn:sdof_region_mac_theorem_1} \\
 d_i &\ge 0, & i=1,\dots,K
\label{eqn:sdof_region_mac_theorem_2}
\end{align}
for almost all channel gains.
\end{theorem}

\begin{theorem}[\!\! \cite{xie_sdof_region,
xie_itw_2014}]
\label{theo:sdofregion:ic_ee_sdof_region}
The \sdof region $D$ of $K$-user IC-EE,  IC-CM, and IC-CM-EE
is the set of all $\mathbf{d}$ satisfying
\begin{align}
 K d_i +  \sum_{j=1,j\neq i}^K d_j &  \le K-1, \quad i=1,\ldots,K
\label{eqn:sdofregion:ic_ee_converse_1}  \\
 \sum_{i\in V} d_i &   \le 1, \quad \forall \ V \subseteq \{1,\ldots,K\}, \ |V|=2
\label{eqn:sdofregion:ic_ee_converse_2} \\
 d_i &  \ge 0,  \quad  i=1,\ldots,K
\label{eqn:sdofregion:ic_ee_converse_3}
\end{align}
for almost all channel gains.
\end{theorem}

The complete proofs can be found in \cite{xie_asilomar_2013,xie_itw_2014, xie_sdof_region}. The major challenge in the proofs of both theorems is to show the tightness of the converse regions. We first note that the converse regions have \emph{polytope} structures. This is because: A set $P \subseteq R^n$ is a \emph{polyhedron} if there is a system of finitely many inequalities $\mathbf{H} \mathbf{x} \le \mathbf{h}$ such that,
\begin{align}
P =\big\{\mathbf{x} \in R^n \,\,|\,\, \mathbf{H} \mathbf{x} \le \mathbf{h}\big\}
\end{align}
Further, if $P$ is a bounded polyhedron, then it is a polytope, which is the case for the converse regions we derive. Due to the Minkowski theorem below, the converse regions are equal to the convex hull of their corresponding extreme points.

\begin{theorem}[{{Minkowski, 1910. \cite[Theorem 2.4.5]{convex_polytopes}}}]
\label{theo:sdofregion:minkowski}
Let $P \subseteq R^n$ be a compact convex set. Then,
\begin{equation}
P = \convhull(\expoints(P)).
\end{equation}
\end{theorem}

Minkowski theorem plays an important role in this problem, since it tells that, instead of studying the polytope $P$ itself, for this problem, i.e., achievability proofs, we can simply concentrate on all extreme points $\expoints(P)$. The following theorem helps us find all extreme points of a polytope $P$ efficiently:  We  select any $n$ linearly independent {active/tight} boundaries and check whether they give a point in the polytope $P$.

\begin{theorem}[{{\!\!\cite[Theorem 7.2(b)]{linear_optimization_and_extension}}}]
\label{theo:sdofregion:polyhedron_ep_rank}
$\mathbf{x}\in R^n$ is an extreme point of polyhedron $P(\mathbf{H}, \mathbf{h})$ if and only if $\mathbf{H} \mathbf{x} \le \mathbf{h}$ and $\mathbf{H}' \mathbf{x} = \mathbf{h}'$ for some $n\times (n+1)$ sub-matrix $(\mathbf{H}',\mathbf{h}')$ of $(\mathbf{H},\mathbf{h})$ with $\rank(\mathbf{H}')=n$.
\end{theorem}

As shown by the proof in \cite{xie_sdof_region}, the \sdof region of the multiple access wiretap channel is constrained by secrecy constraints only. However, different portions of the \sdof region of the interference channel are governed by different upper bounds. To see this, we can study the structure of  the extreme points of $D$, since $D$ is the convex hull of them. The sum \sdof tuple, which is symmetric and has no zero elements, is governed by the upper bounds in \eqn{eqn:sdofregion:ic_ee_converse_1} due to secrecy constraints. However, as shown in \cite{xie_sdof_region}, all other extreme points have zeros as some elements, and therefore are  governed by the upper bounds in \eqn{eqn:sdofregion:ic_ee_converse_2} due to interference  constraints in \cite{multiplexing_gain_of_networks, interference_alignment}. An explanation can be provided as follows: When some transmitters do not have messages to transmit, we may employ them as ``helpers''. Even though secrecy constraint is considered in our problem, with the help of the ``helpers'', the effect due to the existence of the eavesdropper in the network can be \emph{eliminated}. Hence, this portion of the \sdof region is dominated by the interference constraints.

Here, as concrete examples, we provide the \sdof regions for the multiple access wiretap channel and the interference channel with secrecy constraints when $K=2,3,4$, to show intricate differences. The detailed proofs and the structures of the extreme points for all $K$ can be found in \cite{xie_sdof_region}.

\begin{figure}[t]
\centering
\includegraphics[width=0.35\textwidth]{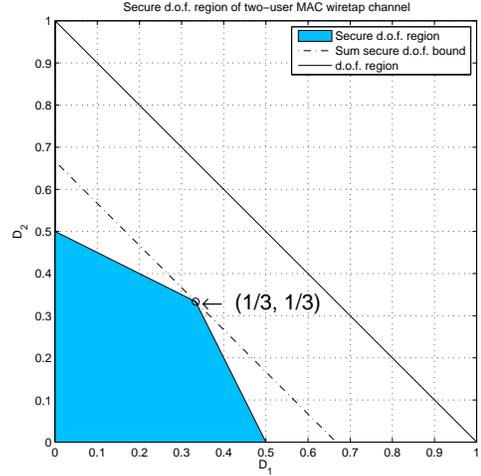}
\caption{The \sdof region of the $K=2$-user multiple access wiretap channel.}
\label{fig:sdof_region_2_user_mac}
\vspace*{-0.4cm}
\end{figure}

For $K=2$, the \sdof region of the multiple access wiretap channel in Theorem~\ref{theo:sdof_mac_capacity_region} becomes
\begin{align}
\label{eqn:sdofregion:mac_region_K_is_two}
D= \Big\{\mathbf{d} : ~  2 d_1 +  d_2 & \le 1,\nl
                          d_1 + 2 d_2 & \le 1, \nl
                              d_1,d_2 & \ge 0\Big\}
\end{align}
and is shown in Fig.~\ref{fig:sdof_region_2_user_mac}. The extreme points of this region are: $(0,0), (\frac{1}{2},0), (0,\frac{1}{2})$, and $(\frac{1}{3}, \frac{1}{3})$. In order to provide the achievability of the region,  it suffices to provide the achievability of these extreme points. In fact, the achievabilities of $(\frac{1}{2},0), (0,\frac{1}{2})$ were proved in \cite{xie_gwch_allerton, xie_sdof_networks_in_prepare} in the helper setting and the achievability of $(\frac{1}{3}, \frac{1}{3})$ was proved in \cite{xie_ulukus_isit_2013_mac, xie_sdof_networks_in_prepare}. Note that $(\frac{1}{3}, \frac{1}{3})$ is the only sum \sdof optimum point.

\begin{figure}[t]
\centerline{\includegraphics[width=0.57\textwidth]{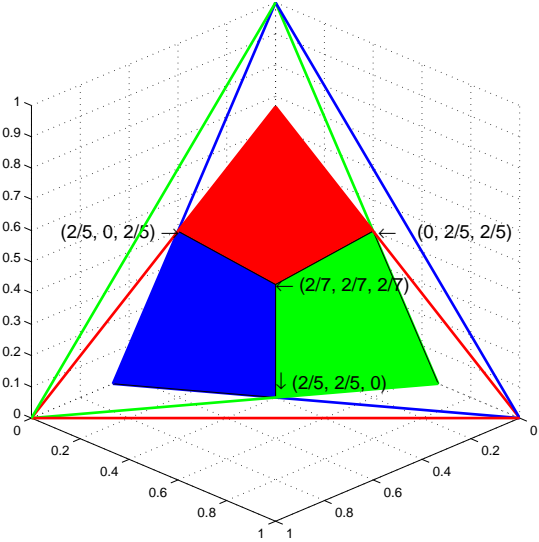}}
\caption{The \sdof region of the $K=3$-user multiple access wiretap channel.}
\label{fig:sdof_region_3_user_mac}
\vspace*{-0.4cm}
\end{figure}

For $K=3$, the \sdof region of the multiple access wiretap channel in Theorem~\ref{theo:sdof_mac_capacity_region} becomes
\begin{align}
  D = \Big\{ \mathbf{d} : ~
    3 d_1 + 2d_2 + 2d_3 & \le 2, \nl
 2 d_1 + 3d_2 + 2d_3  & \le 2, \nl
 2 d_1 + 2d_2 + 3d_3  & \le 2, \nl
        d_1,d_2, d_3  & \ge 0
\Big\}
\end{align}
and is shown in Fig.~\ref{fig:sdof_region_3_user_mac}. The extreme points of this region are:
\begin{align}
\begin{aligned}
& \bigg(0,0,0\bigg)\\
& \bigg(\frac{2}{3}, 0  , 0  \bigg), \bigg(0  , \frac{2}{3}, 0  \bigg), \bigg( 0  , 0, \frac{2}{3}  \bigg)\\
& \bigg(\frac{2}{5}, \frac{2}{5}, 0  \bigg), \bigg(\frac{2}{5}, 0, \frac{2}{5} \bigg), \bigg(0, \frac{2}{5},  \frac{2}{5} \bigg) \\
& \bigg(\frac{2}{7}, \frac{2}{7}, \frac{2}{7} \bigg)
\end{aligned}
\label{eqn:sdofregion:mac_example_K_three}
\end{align}
which correspond to the maximum individual \sdof (see Gaussian wiretap channel with two helpers \cite{xie_gwch_allerton, xie_sdof_networks_in_prepare}), the maximum sum of pair of \sdof (see two-user Gaussian multiple access wiretap channel with one helper, proved in \cite{xie_sdof_region}), and the maximum sum \sdof (see three-user Gaussian multiple access wiretap channel \cite{xie_ulukus_isit_2013_mac, xie_sdof_networks_in_prepare}). Note that $(\frac{2}{7},
\frac{2}{7}, \frac{2}{7})$ is the only sum \sdof optimum point.

For $K=2$, the \sdof region of the interference channel with secrecy constraints in Theorem~\ref{theo:sdofregion:ic_ee_sdof_region} becomes
\begin{align}
\label{eqn:sdofregion:ic_example_k_two}
D= \Big\{\mathbf{d} : ~  2 d_1 +  d_2 & \le 1,\nl
                          d_1 + 2 d_2 & \le 1, \nl
                              d_1,d_2 & \ge 0\Big\}
\end{align}
which is the same as \eqn{eqn:sdofregion:mac_region_K_is_two}, and is shown in Fig.~\ref{fig:sdof_region_2_user_mac}. Note that \eqn{eqn:sdofregion:ic_ee_converse_2} is not necessary for
the two-user case, since summing the bounds $2 d_1 +  d_2  \le 1$ and $ d_1 +  2d_2  \le 1$ up gives a new bound
\begin{align}
  d_1 + d_2 \le \frac{2}{3}
\end{align}
which is the result in Theorem~\ref{eqn:kic-ds-final} and makes the constraint in \eqn{eqn:sdofregion:ic_ee_converse_2} strictly loose. In order to provide the achievability, it suffices to check that the extreme points $(0,0)$, $(\frac{1}{2},0), (0,\frac{1}{2})$, and $(\frac{1}{3}, \frac{1}{3})$
are achievable. In fact, the achievabilities of $(\frac{1}{2},0), (0,\frac{1}{2})$ are similar to \cite{xie_gwch_allerton, xie_sdof_networks_in_prepare} and shown in \cite{xie_sdof_region}. The achievability of $(\frac{1}{3}, \frac{1}{3})$ was proved in \cite{xie_ulukus_isit_2013_kic,xie_unified_kic}. Note that $(\frac{1}{3}, \frac{1}{3})$ is the only sum \sdof optimum point.

\begin{figure}[t]
\centerline{\includegraphics[width=0.40\textwidth]{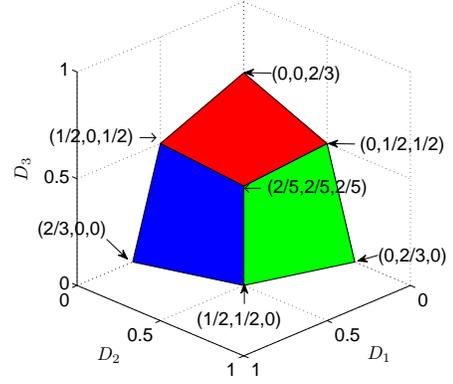}}
\caption{The \sdof region of the $K=3$-user interference channel.}
\label{fig:sdof_region_3_user_ic}
\vspace*{-0.5cm}
\end{figure}

For $K=3$, the \sdof region of the interference channel with secrecy constraints in Theorem~\ref{theo:sdofregion:ic_ee_sdof_region} becomes
\begin{align}
  D = \Big\{ \mathbf{d} : ~
    3 d_1 + d_2 + d_3 & \le 2, \nl
  d_1 + 3d_2 + d_3  & \le 2, \nl
  d_1 + d_2 + 3d_3  & \le 2, \nl
        d_1,d_2, d_3  & \ge 0
\Big\}
\label{eqn:sdofregion:ic_three_user_example}
\end{align}
which is shown in Fig.~\ref{fig:sdof_region_3_user_ic}. Inequality in \eqn{eqn:sdofregion:ic_ee_converse_2} is not necessary for the three-user case, either. This is because, due to the positiveness of each element in $\mathbf{d}$, from the first two inequalities in \eqn{eqn:sdofregion:ic_three_user_example}, we have
\begin{align}
    3 d_1 + d_2 \le 3 d_1 + d_2 + d_3 & \le 2
    \label{eqn:sdofregion:ic_example_k_3_ineqn_1} \\
     d_1 + 3d_2 \le d_1 + 3d_2 + d_3  & \le 2
     \label{eqn:sdofregion:ic_example_k_3_ineqn_2}
\end{align}
Summing the left hand sides up of \eqn{eqn:sdofregion:ic_example_k_3_ineqn_1} and  \eqn{eqn:sdofregion:ic_example_k_3_ineqn_2} gives us
\begin{align}
  d_1 + d_2 \le 1
\end{align}
which is \eqn{eqn:sdofregion:ic_ee_converse_2} with $V=\{1,2\}$, and we have \eqn{eqn:sdofregion:ic_ee_converse_2} for free from \eqn{eqn:sdofregion:ic_three_user_example}. The extreme points of this region are:
\begin{align}
\begin{aligned}
& \bigg(0,0,0\bigg)\\
& \bigg(\frac{2}{3}, 0  , 0  \bigg), \bigg(0  , \frac{2}{3}, 0  \bigg), \bigg( 0  , 0, \frac{2}{3}  \bigg)\\
& \bigg(\frac{1}{2}, \frac{1}{2}, 0  \bigg), \bigg(\frac{1}{2}, 0, \frac{1}{2} \bigg), \bigg(0, \frac{1}{2},  \frac{1}{2} \bigg) \\
& \bigg(\frac{2}{5}, \frac{2}{5}, \frac{2}{5}\bigg)
\end{aligned}
\label{eqn:sdofregion:kic_example_K_three}
\end{align}
which correspond to the maximum individual \sdof (see Gaussian wiretap channel with two helpers \cite{xie_gwch_allerton, xie_sdof_networks_in_prepare}), the maximum sum of pair of \sdof (proved in \cite{xie_sdof_region}), and the maximum sum \sdof (see three-user Gaussian IC-CM-EE in \cite{xie_ulukus_isit_2013_kic, xie_unified_kic}). Note that, $(\frac{1}{2}, \frac{1}{2})$ is the maximum sum \dof for a two-user IC \emph{without} secrecy constraints, and $(\frac{2}{5}, \frac{2}{5}, \frac{2}{5})$ is the only sum \sdof optimum point. Finally, note the difference of the extreme points of the $3$-user interference channel in \eqn{eqn:sdofregion:kic_example_K_three} from the corresponding $3$-user multiple access wiretap channel in \eqn{eqn:sdofregion:mac_example_K_three}, even though the \sdof regions and the extreme points of the $2$-user interference channel and $2$-user multiple access wiretap channel in  \eqn{eqn:sdofregion:ic_example_k_two} and \eqn{eqn:sdofregion:mac_region_K_is_two} were the same.

For $K=4$, the \sdof region of the interference channel with secrecy constraints in Theorem~\ref{theo:sdofregion:ic_ee_sdof_region} becomes
\begin{align}
  D = \Big\{ \mathbf{d} : ~
    4 d_1 + d_2 + d_3 + d_4 & \le 3, \nl
    d_1 + 4 d_2 + d_3 + d_4 & \le 3, \nl
    d_1 + d_2 + 4 d_3 + d_4 & \le 3, \nl
    d_1 + d_2 + d_3 + 4 d_4 & \le 3, \nl
                    d_1+d_2 & \le 1, \nl
                    d_1+d_3 & \le 1, \nl
                    d_1+d_4 & \le 1, \nl
                    d_2+d_3 & \le 1, \nl
                    d_2+d_4 & \le 1, \nl
                    d_3+d_4 & \le 1, \nl
        d_1,d_2, d_3, d_4  & \ge 0
\Big\}
\end{align}
The extreme points of this region are:
\begin{align}
\begin{aligned}
& \bigg(0,0,0,0 \bigg)\\
& \bigg(\frac{3}{4}, 0  , 0, 0  \bigg),
  \bigg(0  , \frac{3}{4}, 0  , 0 \bigg),
  \bigg( 0  , 0, \frac{3}{4} , 0 \bigg),
  \bigg( 0  , 0, 0, \frac{3}{4} \bigg)\\
& \bigg(\frac{2}{3},\frac{1}{3},0,0\bigg) \quad\quad \textrm{up to element reordering}\\
& \bigg(\frac{1}{2}, \frac{1}{2}, \frac{1}{2}, 0  \bigg),
  \bigg(\frac{1}{2}, \frac{1}{2}, 0, \frac{1}{2} \bigg),
  \bigg(\frac{1}{2}, 0, \frac{1}{2}, \frac{1}{2} \bigg) ,
  \bigg(0,\frac{1}{2}, \frac{1}{2}, \frac{1}{2} \bigg)\\
& \bigg(\frac{3}{7}, \frac{3}{7}, \frac{3}{7}, \frac{3}{7} \bigg)
\end{aligned}
\label{eqn:sdofregion:kic_example_K_four}
\end{align}
Here, in contrast to the two-user and three-user cases, \eqn{eqn:sdofregion:ic_ee_converse_2} is absolutely necessary. For example, the point $(\frac{3}{5}, \frac{3}{5}, 0,0)$ satisfies \eqn{eqn:sdofregion:ic_ee_converse_1}, but not \eqn{eqn:sdofregion:ic_ee_converse_2}. In fact, it cannot be achieved, and \eqn{eqn:sdofregion:ic_ee_converse_2} is strictly needed to enforce that fact.

\section{Helper Network with No Eavesdropper CSI: Blind Cooperative Jamming}
In this section, we consider the case where the legitimate transmitters do not have CSI of the channels to the eavesdropper. We present one more technical tool, \emph{blind cooperative jamming}, which will be used to prove that, even in the case of no eavesdropper CSI at the transmitters, the \sdof of the Gaussian wiretap channel with $M$ helpers is still $\frac{M}{M+1}$, as in the case of full eavesdropper CSI in Section~\ref{sec:gwch_m_helper}.

\begin{theorem}[\!\!\cite{xie_blind_cj_ciss_2013}]
The \sdof of the Gaussian wiretap channel with $M$ helpers but no eavesdropper CSI at the transmitters is $\frac{M}{M+1}$ for almost all channel gains.
\end{theorem}
The converse for this result follows from the converse for the case of full CSI, as the \sdof with full CSI is an upper bound for the \sdof without eavesdropper CSI.

\begin{figure*}[t]
\centering
\includegraphics[width=0.8\textwidth]{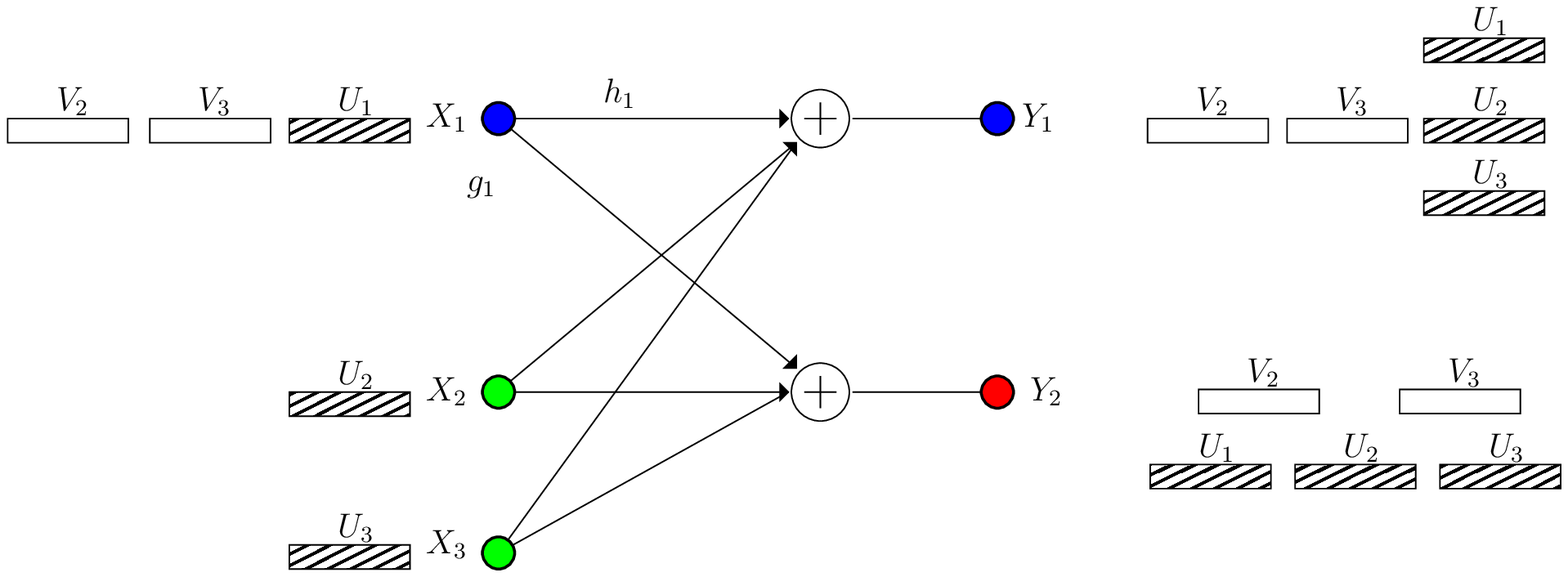}
\caption{Illustration of the alignment scheme based on blind cooperative jamming for Gaussian wiretap
channel with $M$ helpers (eavesdropper's CSI is  not available at the transmitters).}
\label{fig:gwc_no_csi_one_helper_no_csi_ia}
\vspace*{-0.4cm}
\end{figure*}

When there is no eavesdropper CSI at the transmitters, the cooperative jamming signals cannot be aligned with the message carrying signals at the eavesdropper to protect them as in Fig.~\ref{fig:gwc_m_helper_ia}. In this case, the insight of \emph{blind cooperative jamming} is that all of the $M+1$ transmitters send a large number of cooperative jamming signals, which get distributed to sufficiently many dimensions at the eavesdropper's observation space, exceeding its maximum decoding capability and protecting the message carrying signals; see Fig.~\ref{fig:gwc_no_csi_one_helper_no_csi_ia}. Then, the information leakage to the eavesdropper can be upper bounded by a function which vanishes as the transmit power $P$ becomes large, using a method different than in Section~\ref{sec:gwch_m_helper}. In addition, the CSI of the channels to the legitimate receiver is used to align all of the $M+1$ cooperative jamming signals in the smallest possible dimension at the legitimate receiver.

Let $\{V_2,V_3,\cdots,V_{M+1},U_1,U_2,U_3,\cdots,U_{M+1}\}$ be mutually independent discrete random variables, each of which  uniformly drawn from the same PAM constellation $C(a,Q)$ in (\ref{constel}). We choose the input signal of the legitimate transmitter as
\begin{equation}
  X_1  = \frac{1}{h_1}U_1 +  \sum_{k=2}^{M+1} \alpha_k  V_k
\end{equation}
where $\{\alpha_k\}^{M+1}_{k=2}$ are rationally independent and independent of all channel gains. The input signal of the $j$th helper, $j=2,\cdots,M+1$, is chosen as
\begin{equation}
X_j = \frac{1}{h_j} U_j
\end{equation}
Then, the observations of the receivers are
\begin{align}
Y_1
& = \sum_{k=2}^{M+1} {h_1 \alpha_k} V_k
  + \left[ \sum_{j=1}^{M+1} U_j \right]+ N_1 \label{align-no-csi}\\
Y_2
& =
    \sum_{k=2}^{M+1} {g_1 \alpha_k} V_k
    +  \sum_{j=1}^{M+1} \frac{g_j}{h_j} U_j + N_2
\end{align}
where the signals in square parentheses in (\ref{align-no-csi}) are \emph{aligned} at the legitimate receiver.

The intuition here is as follows: We use $M$ independent sub-signals $V_k$, $k=2,\cdots,M+1$, to represent the original message $W$. The input signal $X_1$ is a linear combination of $V_k$s and a cooperative jamming signal $U_1$. At the legitimate receiver, all of the cooperative jamming signals $U_k$s are well-aligned such that they occupy a  small portion of the signal space. Since $\left\{1, h_1 \alpha_2, h_1 \alpha_3, \cdots, h_1 \alpha_{M+1}\right\}$ are rationally independent with probability one, the signals $\left\{V_2,V_3,\cdots,V_{M+1}, \sum_{j=1}^{M+1} U_j \right\}$ can be distinguished by the legitimate receiver. Due to the fact that the eavesdropper's CSI is not available at the transmitters, the alignment-based achievable scheme in Section~\ref{sec:gwch_m_helper} does not work for this model. However, we observe that the coefficients $\left\{\frac{g_1}{h_1}, \cdots, \frac{g_{M+1}}{h_{M+1}}\right\}$ are rationally independent, and therefore, $\left\{U_1,U_2,\cdots,U_{M+1}\right\}$ \emph{span} the \emph{entire space} at the eavesdropper; see Fig.~\ref{fig:gwc_no_csi_one_helper_no_csi_ia}. Here, by \emph{entire space}, we mean the maximum number of \emph{dimensions} eavesdropper is capable to decode, which is $M+1$ in this case. Since the entire space at the eavesdropper is occupied by the cooperative jamming signals, the message signals $\{V_2, V_3, \cdots, V_{M+1}\}$ are protected.

We note that, while only the helpers sent cooperative jamming signals in the case of full eavesdropper CSI in Section~\ref{sec:gwch_m_helper}, here the legitimate transmitter also sends a cooperative jamming signal. These $M+1$ cooperative jamming signals are needed to protect $M$ message carrying signals at the eavesdropper, i.e., the lack of CSI of the eavesdropper is compensated by increasing the number of cooperative jamming signals with respect to the number of message carrying signals.

\section{Conclusions}
In this review paper, we revisited the sum \sdof and \sdof regions of several one-hop wireless networks with secrecy constraints: Gaussian wiretap channel with helpers, Gaussian multiple access wiretap channel, and Gaussian interference channel with secrecy constraints. We first reviewed two key lemmas required for converse proofs. The \emph{secrecy penalty} lemma showed that the net effect of an eavesdropper on the system is that it eliminates one of the independent channel inputs. The \emph{role of a helper} lemma developed a direct relationship between the cooperative jamming signal of a helper and the message rate. We showed how to apply these two lemmas in the helper network in depth, and also in the IC-EE network briefly. We presented achievable schemes based on (asymptotic) real interference alignment, cooperative jamming, structured signalling, and also blind cooperative jamming in the case of no CSI at the transmitters in the helper network. We also reviewed the polytope structure of the \sdof converse regions, identified the extreme points, and then showed the achievability for each of the extreme points.


\begin{IEEEbiography}[{\includegraphics[width=1in, height=1.25in,clip, keepaspectratio]{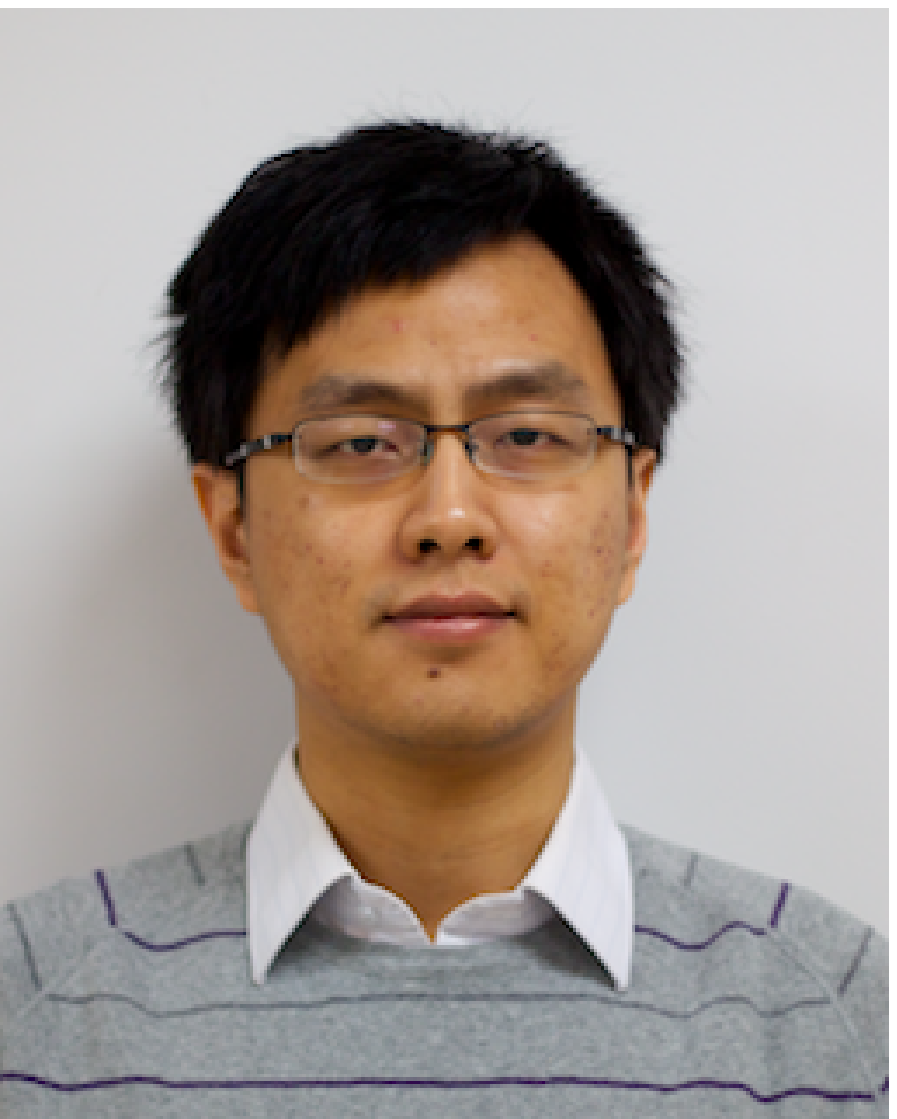}}]{Jianwei Xie} received his Ph.D. degree from the Department of Electrical and
Computer Engineering at the University of Maryland, College Park in May
2014. Prior to that, he received the B.S.~and M.S.~degrees in electronic engineering from the Tsinghua University, Beijing, China, in 2006 and 2008, respectively. Currently, he is with Google Inc., Mountain View, CA USA. 

He received the Distinguished Dissertation Fellowship from the ECE Department at the University of Maryland, College Park, in 2013. His research interests include information theory and wireless communications.
\end{IEEEbiography}

\begin{IEEEbiography}[{\includegraphics[width=1in, height=1.25in,clip, keepaspectratio]{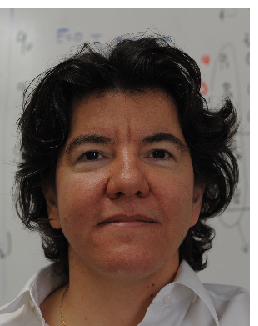}}]{Sennur Ulukus}  is a Professor of Electrical and Computer Engineering at the University of Maryland at College Park, where she also holds a joint appointment with the Institute for Systems Research (ISR). Prior to joining UMD, she was a Senior Technical Staff Member at AT\&T Labs-Research. She received her Ph.D. degree in Electrical and Computer Engineering from Wireless Information Network Laboratory (WINLAB), Rutgers University, and B.S. and M.S. degrees in Electrical and Electronics Engineering from Bilkent University. Her research interests are in wireless communication theory and networking, network information theory for wireless communications, signal processing for wireless communications, information theoretic physical layer security, and energy harvesting communications.

Dr. Ulukus received the 2003 IEEE Marconi Prize Paper Award in Wireless Communications, an 2005 NSF CAREER Award, the 2010-2011 ISR Outstanding Systems Engineering Faculty Award, and the 2012 George Corcoran Education Award. She served as an Associate Editor for the IEEE Transactions on Information Theory (2007-2010) and IEEE Transactions on Communications (2003-2007). She served as a Guest Editor for the IEEE Journal on Selected Areas in Communications for the special issue on wireless communications powered by energy harvesting and wireless energy transfer (2015), Journal of Communications and Networks for the special issue on energy harvesting in wireless networks (2012), IEEE Transactions on Information Theory for the special issue on interference networks (2011), IEEE Journal on Selected Areas in Communications for the special issue on multiuser detection for advanced communication systems and networks (2008). She served as the TPC co-chair of the 2014 IEEE PIMRC, Communication Theory Symposium at 2014 IEEE Globecom, Communication Theory Symposium at 2013 IEEE ICC, Physical-Layer Security Workshop at 2011 IEEE Globecom, Physical-Layer Security Workshop at 2011 IEEE ICC, 2011 Communication Theory Workshop (IEEE CTW), Wireless Communications Symposium at 2010 IEEE ICC, Medium Access Control Track at 2008 IEEE WCNC, and Communication Theory Symposium at 2007 IEEE Globecom. She was the Secretary of the IEEE Communication Theory Technical Committee (CTTC) in 2007-2009.
\end{IEEEbiography}

\end{document}